# Clustered Hierarchy in Sensor Networks: Performance and Security

Mohammed Abuhelaleh, Khaled Elleithy and Thabet Mismar

School of Engineering, University of Bridgeport
Bridgeport, CT 06604
{mabuhela, elleithy, tmismar} @bridgeport.edu

### ABSTRACT

*Many papers have been proposed in order to increase the wireless sensor networks performance; This kind of network has limited resources, where the energy in each sensor came from a small battery that sometime is hard to be replaced or recharged. Transmission energy is the most concern part where the higher energy consumption takes place. Clustered hierarchy has been proposed in many papers; in most cases, it provides the network with better performance than other protocols. In our paper, first we discuss some of techniques, relates to this protocol, that have been proposed for energy efficiency; some of them were proposed to provide the network with more security level. Our proposal then suggests some modifications to some of these techniques to provide the network with more energy saving that should lead to high performance; also we apply our technique on an existing one that proposed to increase the security level of cluster sensor networks.*

## 1. INTRODUCTION

There are many advantages of using sensor networks. They provide dynamic and wireless communication between nodes in a network, which provides more flexible communication. At the same time, sensor networks have some special characteristics compared to traditional networks, which makes it harder to deal with. The most important property that affects this type of networks is the limitation of the resources available, especially the energy.

Wireless Sensor Networks (WSNs) [2] are a special kind of Ad hoc networks that became one of the most interesting areas for researchers. Routing techniques are the most important issue for these networks, where resources are limited. Cluster-based organization has been proposed to provide an efficient way to save energy during communication [3]. In this kind of organization, nodes are organized into clusters. Cluster heads (CHs) pass messages between groups of nodes (group for each CH) and the base station (BS), (Figure1). This organization provides some energy saving which is the main advantage for proposing this organization. Depending on this organization, LEACH (Low Energy Adaptive Clustering Hierarchy) [3] enhanced security, where the CHs are rotating from node to node in the network making it harder for intruders to know the routing elements and attack them. [4]

In this paper, we discuss some existing work of LEACH and we focus on three important criteria; performance, security, and energy consumption. In section two, we discuss the original work of LEACH, and then in the third section we discuss one of the most interesting modifications proposed for LEACH to increase network performance (TCCA). In the fourth and fifth sections, we discuss another two techniques that have been proposed to increase the security of LEACH (F-LEACH and Sec-LEACH). Then, in the sixth and seventh sections, we discuss our proposal and we explain the main modifications that we applied on LEACH and Sec-LEACH to improve network performance and network security. In section 8, we discuss our experiment that we applied to show the improvements that might be gained from applying our protocol compared to the existing protocols.

## 2. LEACH

Low Energy Adaptive Clustering Hierarchy has been presented by [1] to balance the draining of energy during communication between nodes in sensor networks. The BS assumed to be directly reachable by all nodes by transmitting with high enough power. Nodes send their sensor reports to their CHs, which then combine the reports in one aggregated report and send it to the BS. To avoid the energy draining of limited sets of CHs, LEACH rotates CHs randomly among all sensors in the network in order to distribute the energy consumption among all sensors. It works in rounds; in each round, LEACH elects CHs using a distributed algorithm and then dynamically clusters the remaining sensors around the CHs. Sensor-BS communication then uses this clustering result for the rest of the round. (See Fig.1)

### 2.1. LEACH Protocol

Routing in LEACH works in rounds and each round is divided into two phases, the Setup phase and the Steady State phase; each sensor knows when each round starts, using a synchronized clock [1, 2].

Initially, each sensor decides if it will be a CH or not based on the desired percentage of the CHs for the network, and the number of times the sensor has been a CH (to control the energy consumption), this decision is made by the sensor (s) choosing a random number between zero and one. Then it calculates the threshold for itself T(s), then it compares the random number with resulting T(s); if the number is less than T(s), the sensor becomes a CH for the current round. T(s) for x round with desired percentage of cluster heads P is calculated by:

$$T(s) = \begin{cases} \dfrac{P}{1 - P * (x \bmod \dfrac{1}{P})} \ldots\ldots if n \in G \\ 0 \ldots\ldots\ldots\ldots\ldots\ldots\ldots\ldots\ldots otherwise \end{cases} \qquad (1)$$

Where G is a set of nodes that have not been CHs in the last 1/p round.

Setup phase includes three steps. Step1 is the advertisement step, where each sensor decides its probability to become a CH, based on the desired percentage of CHs and its remaining energy, for the current round; a sensor who decides to become a CH broadcasts an advertising message to other nodes that it is ready to become a CH. Carrier sense multiple access protocol is used to avoid collisions. Cluster joining step is the second step, where the remaining sensors pick a cluster to join according to the highest signal received; then they send request messages to the desired CHs. Step three starts after the CHs receive all requests from other sensors, where CHs broadcast confirmation messages to their cluster members; these messages include the time slot schedule to be used during the steady state phase.

The Steady State phase (the actual communication) then starts. It consists of two steps; in the first step each node starts by sending its sensor report to its CH based on the time provided by the time slot schedule. When a CH receives all the reports, it aggregates them in one report and it sends this report to the BS (step 2). Next we show the details of each step by providing the content of each one; for this purpose we combine the two phases in one phase with five steps.

In step one, a CH broadcasts to the rest of sensors, its ID and the Advertising message, then, in step two, each sensor sends its ID, CH ID, and the Join Request message to its desired CH. When a CH receives all requests, it broadcasts its ID, and the time slot schedule for sensors that includes each member with its time slot (step three). Each sensor then sends its ID, CH ID, and the sensing report to its CH (step four). Finally, each CH sends its ID, BS ID, and the aggregate report of its members to the BS.

The transmission of information between sensors, and between sensors and BSs, are performed using CSMA MAC protocol. On the other hand, they communicate using CDMA codes to reduce the interference that may occur from communication of nearby nodes.

## 2.2. Energy saving in LEACH

LEACH provides many techniques for saving the total energy during network communication; in this section we focus on the main concepts that have been applied by LEACH to save the energy.

LEACH is a self-organization adaptive protocol, and it uses randomization to evenly distribute the energy load among the sensors in the network; this and the random way that CHs rotate around the various sensors reduce the possible draining of the battery for each sensor.

A local data compression to compress the amount of data being sent from clusters to BS is used to reduce the energy consumption and to enhance the system lifetime.

The time schedule that is being sent by CHs to their members, gives break time for sensors that have not reached their time yet to be in sleeping mode, which helps them save their energy for their scheduled time.

Finally, the nature of the way that LEACH changes CHs each round, and the way that each CH can be elected, provides high energy saving for the whole network.

## 2.3. Security in LEACH

LEACH is more powerful against attacks than most other routing protocols [2, 4]. CHs in LEACH that directly communicate with BS can be anywhere in the network and they are changing from round to round, which makes it harder for intruders to identify the critical nodes in the network.

On the other hand, LEACH is vulnerable to a number of security attacks [2, 4], including spoofing, jamming, and replay attacks. Since LEACH is a cluster based protocol, it relies mainly on the CHs for routing and data aggregation, which makes the attacks involving CHs, the most harmful attacks.

Some kinds of attacks, such as sinkhole and selective forwarding, may occur if an intruder manages to become a CH, which results in disrupting the work of the network.

# 3. TCCA

Time-Controlled Clustering Algorithm (TCCA) allows multi-hop clusters using message time-to-live (TTL) and timestamp to control the way the clusters form. Residual energy is also considered before a sensor volunteers to become a CH, and a numerical model is provided to quantify its efficiency on energy usage.

## 3.1. TCCA Protocol

Similar to LEACH, TCCA's operation is divided into rounds with two phases concluded in each round (Setup phase and the Steady State phase). CHs are elected and the clusters are formed in Setup phase; then the complete cycle of data collection, aggregation and transfer to the BS occurs in the Steady State phase.

To determine the eligibility of sensor to be CH, TCCA adds some modifications to the LEACH technique. A sensor residual energy is considered and a random number between 0 and 1 (Tmin) is generated by each sensor to determine its eligibility to become CH. If this number is less than the variable threshold, the sensor becomes a CH for the current round. The threshold for sensor 's' in round r, with desired CH percentage $p$, residential energy RE and maximum energy MaxE is calculated by (2):

$$T(s) = \begin{cases} \max(\dfrac{p}{1 - p(r \bmod \dfrac{1}{p})} \times \dfrac{RE}{MaxE}, T\min)...\forall s \in G \\ \\ 0...\forall s \notin G \end{cases} \quad (2)$$

G is a set of nodes that have not been CHs in the last 1/p round

When CH is elected, it advertises to other sensors to become its members; this advertisement message contains CH ID, initial TTL, timestamp and its residual energy. Sensors receive the message will forward it to their neighbors based on TTL value which may be based on the current energy level of CH; at the same time they join this CH with the rest of sensors who received the message. Once a sensor decides to join the cluster, it informs the corresponding CH by sending a join request message that carries sensor ID, CH ID, the original timestamp from advertising message and the remaining TTL value. The CH uses the timestamp to approximate the relative distance of its neighbors and to learn the best setup phase time for future rounds [3].

## 3.2. Energy saving in TCCA

TCCA applies a new condition for electing CHs by considering the remaining energy of the sensors. At the same time, it guarantees that every sensor will become a CH at least one time per 1/P rounds, where P is the desired percentage of CHs. These modifications provide the network with high energy balance by distributing the energy among all sensors.

TCCA provides optimum cluster size (K) for K-hops in order to produce high performance similar to the performance in the one-hop network. Also it reduces the complexity of transmission schedule generation to O (1).

TCCA uses timestamp and Time to live (TTL) tags to control the cluster formation; this leads to gain more energy balance.

## 3.3. Security in TCCA

TCCA follows the main steps provided by LEACH with some modifications that do not affect the level of security that is provided by LEACH; this means that TCCA does not have enough protection against Spoofing, Jamming, Replay and some other kinds of attacks.

## 4. SEC-LEACH

It proposes a new modification for LEACH to increase the level of security and to protect the network from many kinds of attacks, specially the sinkhole and selective forwarding attacks [2, 4].

Sec-LEACH proposes generating a large pool of keys and their IDs at the time the network is deployed. Each sensor then assigned with a ring of keys taken from key pool pseudo randomly [5]. First it generates a unique ID for each sensor using a pseudo random function (PRF), then a large enough number of keys is assigned to each sensor from the key pool; also assign each node by a pair-wise key shared with the BS [2,5].

## 4.1. Sec-LEACH Protocol

When elected a CH broadcasts its advertising message, it includes the ID of the keys in its key ring, the other sensors cluster around the nearest CH with whom they share a key [2]. The details of Sec-LEACH protocol works as follows:

CHs are elected as in LEACH, and then these CHs broadcast their IDs and their nonce (step 1). In step 2, other sensors computes the set of CHs keys IDs and choose the nearest CH with whom they share a key; these sensors then send Join Request messages, protected by MAC that is produced by the share key, and the nonce that is broadcasted by the CH, to prevent reply attacks; the ID of the key chosen to protect the link is also sent with the to make CH knows which key to use for verifying the MAC. To complete the setup phase, CHs send the time schedule to sensors that choose to become their members (step3).

Step4 is the first step in Steady State phase, where sensor-to-CH communications are protected using the same key used to protect the Join Request message. To prevent replay, a value computed from the nonce and the reporting cycle is also included. The CH then decrypts sensor reports and performs a data aggregation then sends it to the BS protected by the symmetric key shared with the BS. A counter is included in the MAC value also, to provide freshness.

## 4.2. Energy saving in Sec-LEACH

Sec-LEACH works like the original LEACH, but here some extra bits have to be added to the total transactions that occur during the communication in the network. As discussed in [2], these overloads will not affect the efficiency of the original LEACH if suitable size of the key pool and suitable number of keys assigned to each sensor are chosen.

## 4.3. Security in Sec-LEACH

Sec-LEACH provides more protection to the network than it is in LEACH, where it protects against spoofing, jamming, and replay attacks. In addition, it prevents sinkhole and selective forwarding attacks.

## 5. F-LEACH

F-LEACH [6] is an enhanced version of LEACH that gives protection for the network. It suggests that each node has to have two symmetric keys: a pair-wise key shared with the BS and a second key chain held by the BS. According to that, it suggested small modifications for LEACH. For the setup phase, the message sent by RNs should consist of an encrypted message that contains the ID of the node that should receive the message and the ID of the CH itself as plain text, and the encryption (ID of CH, the counter shred by CH and the BS, and the advertisement message) using the message authentication code (MAC) that is produced using the shared key between CH and the BS.

It is already proved by [2], that Sec-LEACH provides more security and performance than in F-LEACH, so we will not go through that technique in detail.

# 6. MODIFIED LEACH

To improve LEACH we insinuate Modified LEACH (Mod-LEACH). Mod-LEACH works in two rounds: a Full transmission round and a half transmission one.

Each sensor checks its ability to become a CH depending on the desired percentage of CHs, current round, and the remaining energy; we used the same formula used by TCCA to calculate the threshold.

The sensors that are able to become a CH (ready sensors) for the current round start listening for any query that might be sent by other sensors; the other sensors start broadcasting their reports to their neighbors, the packets contain some other tags to determine the status of the packets; any ready sensor that receives the report saves it temporarily and sends a confirmation/request to the related sensor confirming that it is ready to send its report and providing its ability status to become a CH for next round. Sensors who receive the confirmation, reply back to the CH with another confirmation and save the CH id to use it for the next round (if the status of the CH shows that it is able to be a CH for two rounds). CHs will then collect all the reports that have been confirmed in one compressed report and forwards it to the BS (This is considered as a full transmission round).

For the next round, the sensors with no CHs will repeat the same scenario, and the sensors with CHs will send the report only to their CHs; when the old CHs receive the reports, it will aggregate them in one report and forward it to the BS (this considered as a half transmission report).

Next we will explain in details the complete protocol.

## 6.1. Mod-LEACH Protocol

The operation of the Mod-LEACH occurs in rounds, and rounds are classified into two kinds, the full transmission round and the half transmission round. The main idea here is to skip the setup phase that is proposed by all other discussed protocols.

At the beginning of each round, CHs elect themselves. In order to determine the eligibility of a sensor to be a CH, each sensor (S) generates a random number between 0 and 1; then this number is compared to a sensor variable threshold value T(S); if the value of the threshold is greater than the random number, the sensor becomes a CH for the current round (R). The Threshold value can be calculated using the same formula that is used by TCCA [3]; first it calculates the threshold for two rounds as follows:

$$T(s)a = \begin{cases} \max(\dfrac{P}{1 - P(R \bmod \dfrac{1}{P})} \times \dfrac{\mathrm{Re}\,mEng}{MaxEng * 2}, ...T\min)...ifs \in G \\ 0......................................................, otherwise \end{cases} ...(3)$$

If formula (3) is approved, then it is ready to become a CH for two rounds. If formula (3) is not approved, the sensor will calculate formula (4) to see if it is able to become a CH for only one round.

$$T(s)a = \begin{cases} \max(\dfrac{P}{1 - P(R \bmod \dfrac{1}{P})} \times \dfrac{\mathrm{Re}\,mEng}{MaxEng}, ...T\min)...ifs \in G \\ 0......................................................, otherwise \end{cases} ...(4)$$

Where P is the desired percentage of CHs, Tmin is a minimum threshold (to avoid the possibility of remaining energy shortage), and G is the set of sensors that have not became CHs in 1/P round, MaxEng is the maximum energy that the sensor could have, RemEng is the sensor remaining energy.

Each elected CH starts listening to the network; other sensors start broadcasting their reports to their neighbors (using Carrier sense multiple access protocol for transmission to avoid collisions); this message consists of Sensor ID, report, Requesting type tag (RT: 0 for request, 1 for approves), Time to live (TTL: set to 1, broadcast to only direct neighbors), packet request status tag (PR: 0 for the first packet, 1 for the second packet). When ready sensors (CHs) receive the messages, it saves each report with the node id temporarily in its memory, and then it sends requests with confirmation to those sensors indicating that it is ready to become their CH for the current round, when formula3 applies, it also indicates that it is also ready to be their CH for the next round; the message contains: CH id, pairs of Sensor id with its time (to prevent collisions and provide less delay), TTL (set to 1), RT (set to 1), PR (set to 0) and the ability tag (AT: 0 for one round ability, and 1 for two rounds ability). Sensors receive the message from CHs; if they receive more than one request then they will choose the one with the ability to become CH for two rounds, AT=1 (here it will save the CH id to use it in the next round); if they receive many requests with the same values, then they pick the CH randomly; Sensors then reply to CHs with confirmation; the message contains: Sensor id, CH id, and an Acknowledgment tag (ACK: set to 1). When a CH receives the confirmations it combines all collected reports in one compressed report and forwards it to the BS; the message contains: CH id, BS id and the aggregation report.

In the next round, sensors check first if they are group members of a CH with an ability to handle two rounds, if they are, then they use it for the current round (half transmission round is applied); the sensor sends its report to its CH; the message contains: Sensor id, CH id, PR (set to 1), TTL (set to 1), PR (set to 1). CH receives the reports, aggregate then in one report, and then send them to the BS, the message contains: Ch id, BS id, and the aggregation report; then the CH will send acknowledgments to its members and remove them from its memory; the acknowledgment message contains: Ch id, Sensor id, and ACK (set to 1); sensors who receive the acknowledgment then remove CH info from their memories.

In the case that the sensor does not have a CH from the previous round, it will repeat the first scenario for full round transmission.

## 6.2. Energy saving in Mod-LEACH

Mod-LEACH applies the same condition that has been applied by TCCA for electing CHs by considering the remaining energy of the sensors. At the same time, it guarantees that every sensor will become a CH at least one time per 1/P rounds, where P is the desired percentage of CHs. These modifications provide the network with high energy balance by distributing the energy among all sensors.

Mod-LEACH provides enhanced energy saving by dealing with double round technique, where it saves almost half of the energy used in one regular round; for the full transmission round, it will consume more energy than LEACH and TCCA, but it covers that gab in the next round, and even saves more energy than other protocols may save.

Mod-LEACH uses Time to live (TTL) tags to control the cluster formation, where the broadcasting occurs only on the direct neighbors; this leads to a more energy balanced network.

## 6.3. Security in Mod-LEACH

Mod-LEACH provides the same level of security that has been provided by LEACH and TCCA, where it didn't affect the main idea of these protocols which is the dynamic rotation of CHs around the network.

# 7. MODIFIED SEC-LEACH

Here we apply the same technique that we proposed for Mod-LEACH, in a way that deals with the security proposal that been suggested by Sec-LEACH.

After generating the key pool and assigning the groups of keys for each node, in addition to a pair-wise key for each of them, we can start our protocol.

The sensors start sending their reports to their direct neighbors; the sensors that are ready to become CHs, receive the reports, store them temporarily and inform the desired sensors with their ability to send their reports; the message will also include the duration that the CH may handle the cluster (one or two rounds). The sensors that received the confirmation, replies to the nearest neighbors (higher signals) with higher ability (two rounds) with acknowledgment and a permission to send; the CHs then will combine the reports in one compressed report and forward it to the BS; if the sensor connects with a CH that can handle two consecutive rounds, then it will implicitly consider it as its CH, so in the next round it will only send its report to its CH and then the CH will combine the sensor report with other reports like before and send it to the BS; then the CH will send an acknowledgment to its members that the reports have been sent.

All the steps pass through the encryption technique provided by Sec-LEACH using the random key distribution.

## 7.1. Mod-Sec-LEACH Protocol

The operation of the Mod-Sec-LEACH occurs in rounds, and rounds are classified into two kinds, the full transmission round and the half transmission round. The main idea here is to skip the setup phase that is proposed by all other discussed protocols.

Prior to network deployment, each sensor is assigned a group of keys randomly from big key pool provided by the BS. A pseudo random function is used to produce the keys IDs; These IDs then, map to their corresponding values in the key pool. Also each sensor assigned by a pair-wise key sharing it with the BS for secure direct communication.

At the beginning of each round, CHs elect themselves. In order to determine the eligibility of a sensor to be a CH, each sensor (S) generates a random number between 0 and 1; then this number is compared to a sensor variable threshold value T(S); if the value of the threshold is greater than the random number, the sensor becomes a CH for the current round (R). The Threshold value can be calculated using the same formula that is used by TCCA; first it calculates the threshold for two rounds as in formula (3).

If formula (3) is approved, then it is ready to become a CH for two rounds; if not, then sensor checks its ability to become CH for only one round, by applying formula (4).

Each elected CH starts listening to the network; other sensors start broadcasting their reports to their neighbors (using carrier sense multiple access protocol for transmission to avoid collisions); this message consists of Sensor ID, the ID of the common key that is picked from the sensor keys ring (CK), Requesting type tag (RT: 0 for request, 1 for approves), Time to live (TTL: set to 1, broadcast only to direct neighbors), packet request status tag (PR: 0 for the first packet, 1 for the second packet), and the encryption of sensor id and the report, all encrypted using the Message Authentication Code (MAC) that was produced using CK. When ready sensors (CHs) receive the messages, it checks if it shares the same key/s with those who sent the messages; if it is, then it saves each report with the node id temporarily in its memory, and then it sends requests with confirmation to those sensors indicating that it is ready to become their CH for the current round, when formula5 applies, it also indicates that it is also ready to be their CH for the next round; the message contains: CH id, pairs of Sensor id with its time (to prevent collisions and provide less delay), TTL (set to 1), RT (set to 1), PR (set to 0) and the ability tag (AT: 0 for one round ability, and 1 for two rounds ability). Sensors receive the message from CHs; if they receive more than one request then they will choose the one with the ability to become CH for two rounds, AT=1 (here it will save the CH id to use it in the next round); if they receive many requests with the same values, then they pick the CH randomly; Sensors then reply to CHs with confirmation; the message contains: Sensor id, CH id, CK, an Acknowledgment tag (ACK: set to 1), and the encryption of (Sensor id, CH id, CK, and ACK), encrypted using the MAC produced by sensors CKs. When a CH receives the confirmations it decrypt and combines all the reports that it has in one compressed report and forwards it to the BS; the message contains: CH id, BS id, the aggregation report, and the encryption of (CH id, BS id, and the aggregation report) encrypted by the MAC produced by CH CK.

In the next round, sensors check first if they are group members of a CH with an ability to handle two rounds, if they are, then they use it for the current round (half transmission round is applied); the sensor sends its report to its CH; the message contains: Sensor id, CH id, CK, PR (set to 1), TTL (set to 1), PR (set to 1), and the encryption of (Sensor id, CH id, CK, PR, TTL, and PR), all encrypted using sensor CK . CH receives the reports, decrypt and aggregate them then in one report, and then send them to the BS, the message contains: CH id, BS id, CK, the aggregation report, and the encryption of (CH id, BS id, CK, and the aggregation report) encrypted by the MAC produced from CH CK; then the CH will send acknowledgments to its members and remove them from its memory; the acknowledgment message contains: CH id, Sensor id, CK, ACK (set to 1), and the encryption of (CH id, Sensor id, CK, and ACK) encrypted using the MAC produced from CH CK; sensors that receive the acknowledgment remove CH info from their memories.

In the case that the sensor does not have a CH from the previous round, it will repeat the first scenario for the full round transmission Like in LEACH.

## 8. EXPERIMENTATION AND ANALYSIS

In this section, we discuss the numerical experimentation; here we describe the chosen parameters groups for each protocol, following the same scenario. The experiment is applied on LEACH, TCCA, Sec-LEACH, Mod-LEACH, and Mod-Sec-LEACH protocols; we applied them on three different network sizes (100, 1000, and 10000 sensors); for each size, 1000 rounds were processed with the following initial values of main parameters:
- The desired percentage of CHs (P) is set to 0.05.
- Each sensor starts with 0.5 j energy.
- The amplifier energy is assumed to be 100 pj.
- The electronic energy is assumed to be 50 nj.
- Each sensor data range is set to 30m.
- The message size of a sensor data is set to 50 bits.
- Each node has 2000-bit data packet to send to the BS.

### 8.1. Energy Saving

LEACH provides many techniques to save energy during network communication; where it is a self-organization, adaptive protocol and it uses randomization to evenly distribute the energy load among the sensors in the network, in addition to the random way that CHs rotate around the various sensors which is reducing the possible draining of the battery for each sensor. Also, performing a local data compression to compress the amount of data being sent from clusters to BS reduces the energy consumption and enhances the system lifetime. These factors, in addition to the way that CHs change every cycle provide LEACH with High energy saving. Mod-LEACH applies the same

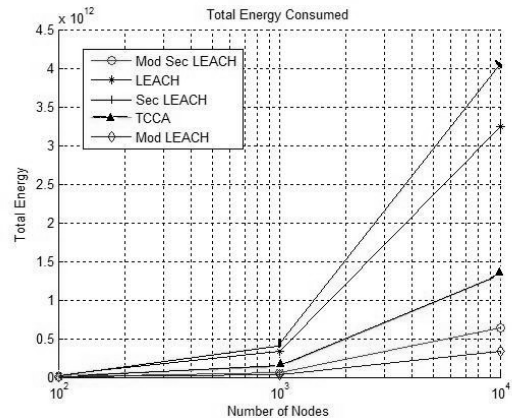

Fig.3. Total energy consumption occur in LEACH, TCCA,
Sec-LEACH, Mod-LEACH and Mod-Sec-LEACH after
1000 rounds, for different network sizes (100, 1000,
10000).

factors to Sensor networks which provide it with similar energy saving to the LEACH at this point.

TCCA adds additional factors to save energy; it uses Time to live (TTL) tag to control the cluster formation, which leads to gain more energy balance. It also uses the new condition provided by TCCA to elect CHs each round, which results in more energy control. [3] Shows that TCCA works almost three times better than LEACH in energy saving. Mod-LEACH and Mod-Sec-LEACH applied TCCA factors, which means that it works three times better than LEACH in energy saving. Now by applying the new idea that we discussed before, we can notice that Mod-LEACH and Mod-Sec-LEACH provide the network with almost four times more energy saving than what is provided by LEACH, and almost double of that in TCCA.

Our experiment shows that the variation of energy consumption is very small when network size is small (i.e. 100 sensors), but it varies more if we increase the network size. Fig.3 shows that, for network size of 10,000 sensors, total energy consumption is minimum in Mod-LEACH with almost $0.3 \times 10^{12}$ nj, then Mod-Sec-LEACH comes with energy consumption of almost $0.7 \times 10^{12}$ nj, then TCCA comes with energy consumption of almost $1.3 \times 10^{12}$ nj at third place, at fourth place LEACH comes with $3.3 \times 10^{12}$ nj, and finally Sec-LEACH comes with $4.1 \times 10^{12}$ nj. The variation comes from the nature of how Mod-LEACH and Mod-Sec-LEACH works; using TTL, in addition to continuous checking of residual energy of each sensor, gives Mod-LEACH and TCCA protocols more energy balance for large network size; double round technique provides Mod-LEACH and Mod-Sec-LEACH with more energy saving.

## 8.2. Data Overload

TCCA works with multi-hops clusters; this reduces the number of clusters, which reduces the total transactions required in network communications; this leads to highly reduce data overload compared with LEACH. Mod-LEACH and Mod-Sec-LEACH have two different round types; in the full transmission round it will produce more data overload than that produced by TCCA,

LEACH, and Sec-LEACH, but by applying the half transmission technique on the next round, we balance the increase in the data in the previous round and we provide less total data overload than that provided in double rounds with LEACH, TCCA, and Sec-LEACH.

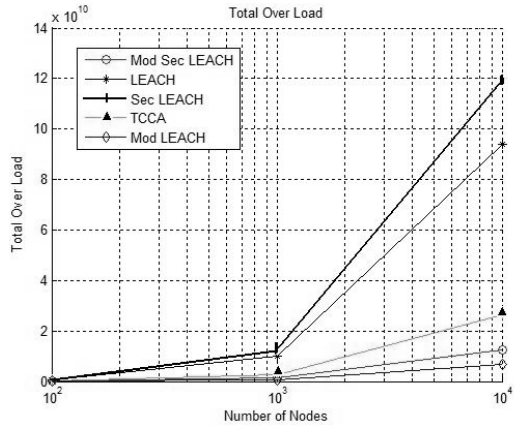

Fig.4. Total data overload in LEACH, TCCA, Sec-LEACH, Mod-LEACH, and Mod-Sec-LEACH after 1000 rounds for different network sizes (100, 1000, 10000).

Our experiment shows that, for a large network size (i.e. 10000 sensors); the total data overload is minimized using Mod-LEACH. Fig.4 shows that with Mod-LEACH data overload reaches almost $0.2 \times 10^{10}$ bits; in Mod-Sec-LEACH, it reaches $0.4 \times 10^{10}$ bits where in TCCA it reaches $2.2 \times 10^{10}$ bits,; in LEACH it reaches $9.3 \times 10^{10}$, and in Sec-LEACH, it reaches $12 \times 10^{10}$ bits; this shows that Sec-LEACH produces data overload almost 12 times than Mod-LEACH and Mod-Sec-LEACH; LEACH produces more data overload, almost nine times more than the data overload produced by Mod-LEACH and Mod-Sec-LEACH. Moreover, TCCA produces more data overload, almost five times than data overload produced by Mod-LEACH and Mod-Sec-LEACH.

### 8.3. Performance

According to the energy saving analysis, we can figure out that the number of Dead Nodes that may appear in LEACH and Sec-LEACH will be much higher than the number of dead nodes in Mod-LEACH and Mod-Sec-LEACH, where the number of Dead Nodes depends on the energy consumption by the network (Fig.5).

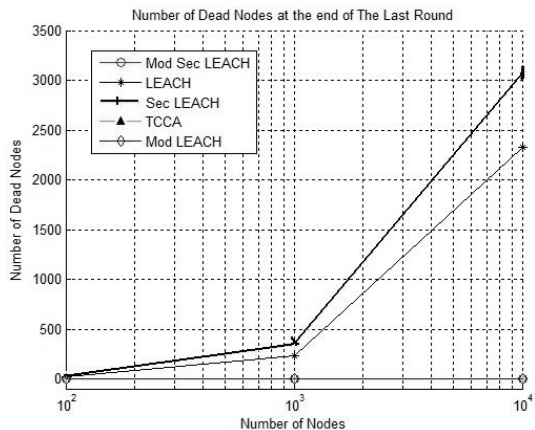

Fig.5. Dead nodes occur in LEACH, TCCA, Sec-LEACH, Mod-LEACH, and Mod-Sec-LEACH after 1000 rounds, for different network sizes (100, 1000, 10000).

## 9. CONCLUSIONS

The results shows that our proposal should provide wireless sensor networks with high performance; also it shows that our proposal is flexible and can be applied to other techniques that are related to the clustered hierarchy without affecting the main purpose of the original technique, more specifically is the security level that can be achieved using that protocol; at the same time it provides it with better performance.

## REFERENCES

[1]   W. R. Heinzelman, A. Chandrakasan, and H. Balakrishnan. Energy-efficient communication protocol for wireless microsensor networks. In IEEE Hawaii Int. Conf. on System Sciences, pages 4–7, January 2000.

[2]   Leonardo B. Oliveira, Hao C. Wong, M. Bern, Ricardo Dahab, A. A. F. Loureiro. SecLEACH - A Random Key Distribution Solution for Securing Clustered Sensor Networks. Fifth IEEE International Symposium on Network Computing and Applications (NCA'06)

[3]   S. Selvakennedy, and S. Sinnappan. A Configurable Time-Controlled Clustering Algorithm for Wireless Networks. 2005 11th International Conference on Parallel and Distributed Systems (ICPADS'05).

[4]   Chris Karlof, Naveen Sastry, and David Wagner. TinySec: A Link Layer Security Architecture for Wireless Sensor Networks. 2004 Conference on Embedded Networked Sensor Systems Proceedings of the 2nd international conference on Embedded networked sensor systems.

[5]   Sencun Zhu, Shouhuai Xu, Sanjeev Setia, and Sushil Jajodia. Establishing Pairwise Keys for Secure Communication in Ad Hoc Networks: A Probabilistic Approach. Proceedings of the 11th IEEE International Conference on Network Prot1ocols (ICNP'03)

### Authors

Dr. Khaled Elleithy received the B.Sc. degree in computer science and automatic control from Alexandria University in 1983, the MS Degree in computer networks from the same university in 1986, and the MS and Ph.D. degrees in computer science from The Center for Advanced Computer Studies at the University of Louisiana at Lafayette in 1988 and 1990, respectively. From 1983 to 1986, he was with the Computer Science Department, Alexandria University, Egypt, as a lecturer. From September 1990 to May 1995 he worked as an assistant professor at the Department of Computer Engineering, King Fahd University of Petroleum and Minerals, Dhahran, Saudi Arabia. From May 1995 to December 2000, he has worked as an Associate Professor in the same department. In January 2000, Dr. Elleithy has joined the Department of Computer Science and Engineering in University of Bridgeport as an associate professor. In May 2003 Dr. Elleithy was promoted to full professor. In March 2006, Professor Elleithy was appointed Associate Dean for Graduate Programs in the School of Engineering at the University of Bridgeport.

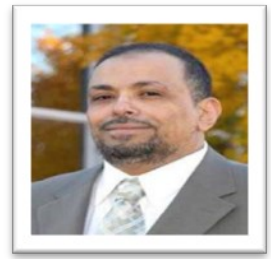

**Mohammed Abuhelaleh** is a full-time Ph.D. student of Computer Science and Engineering at the University of Bridgeport. He worked as a lecturer for Alhusein Bin Talal Universtity; He thought some computer science courses, in addition to college courses, like Data Structure, C++, and Computer Skills for three years. He has master degree Computer Science from University of Bridgeport, and graduated with a GPA of 3.48. Mohammed now is at the end Of fifth semester of PHD program. Mohammed worked as Graduate Assistant for many times under Prof. Elleithy.

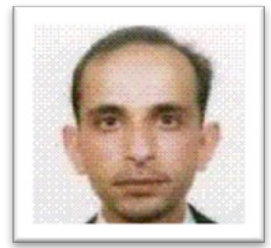

**Thabet Mismar** is a full-time M.Sc. student of Electrical Engineering at the University of Bridgeport. He has B.Sc. degree of Electrical Engineering from the University of Jordan. Thabet is now in the last semester of the M.Sc. program and he worked as a graduate assistant for Prof. Elleithy at the dean of engineering and technology office at the University of Bridgeport.

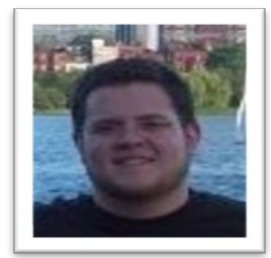